\documentclass{emulateapj}
\usepackage{ulem}
\usepackage{longtable}
\usepackage{graphicx}
\input{epsf.sty}
\input{psfig.sty}

\shorttitle{Baryon Loading}

\begin{document}
\title{Hyper-accreting black hole as GRB central engine. I: Baryon loading in GRB jets}
\author{Wei-Hua Lei\altaffilmark{1,2}, Bing Zhang\altaffilmark{3,2}, En-Wei Liang\altaffilmark{4,5} }
\altaffiltext{1}{School of Physics, Huazhong University of Science and Technology, Wuhan, 430074, China. Email: leiwh@hust.edu.cn}
\altaffiltext{2}{Department of Physics and Astronomy, University of Nevada Las Vegas, 4505 Maryland Parkway, Box 454002, Las Vegas, NV 89154-4002, USA. Email: zhang@physics.unlv.edu}
\altaffiltext{3}{The Kavli Institute for Astronomy and Astrophysics and Department of Astronomy
, Peking University, Beijing 100871, China}

\altaffiltext{4}{Department of Physics and GXU-NAOC Center for Astrophysics and Space Sciences, Guangxi University, Nanning 530004, China}

\altaffiltext{5}{The National Astronomical Observatories, Chinese Academy of Sciences, Beijing 100012, China}

\begin{abstract}
A hyper-accreting stellar-mass black hole has been long speculated as the best candidate of central engine of gamma-ray bursts (GRBs). Recent rich observations of GRBs by space missions such as Swift and Fermi pose new constraints on GRB central engine models. In this paper, we study the baryon loading processes of a GRB jet launched from a black hole central engine. We consider a relativistic jet powered by $\nu \bar\nu$-annihilation or by the Blandford-Znajek (BZ) mechanism. We consider baryon loading from a neutrino-driven wind launched from a neutrino-cooling-dominated accretion flow. For a magnetically dominated BZ jet, we consider neutron-drifting from the magnetic wall surrounding the jet and subsequent positron capture and proton-neutron inelastic collisions. The minimum baryon loads in both types of jet are calculated. We find that in both cases, a more luminous jet tends to be more baryon poor. A neutrino-driven ``fireball'' is typically ``dirtier'' than a magnetically dominated jet, while a magnetically dominated jet can be much cleaner. Both models have the right scaling to interpret the empirical $\Gamma-L_{\rm iso}$ relation discovered recently. Since some neutrino-driven jets have too much baryon loading as compared with the data, we suggest that at least a good fraction of GRBs should have a magnetically dominated central engine. 
\end{abstract}
\keywords{accretion, accretion disks--black hole physics--magnetic field}

\section{Introduction}

Recent observations of gamma-ray bursts (GRBs) with the space missions such as {\em Swift} and {\em Fermi} have greatly enriched our knowledge of this phenomenon. These new data place important constraints on the GRB central engine models. In general, a GRB central engine should meet the following criteria: (1) It must be able to power a very powerful outflow with isotropic luminosity at least that of the gamma-ray luminosity, i.e. $L_{\rm iso} \sim (10^{49}-10^{53})~{\rm erg~s^{-1}}$ (e.g. Zhang \& M\'esz\'aros 2004). (2) The jet must contain a small baryon contamination, so that the it can reach a high Lorentz factor, typically $\Gamma > 100$ (e.g. Lithwick \& Sari 2001). (3) The central engine must be intermittent to power rapid variability as observed in many GRBs (e.g. Fishman \& Meagan 1995). (4) Since a good fraction of GRBs are followed by erratic X-ray flares, the GRB central engine must be long-lived and can power delayed activities (e.g. Burrows et al. 2005; Zhang et al. 2006). (5) In some GRBs (e.g. GRB 080916C), the broad-band spectra show no evidence of quasi-thermal emission from a fireball photosphere (Abdo et al. 2009), suggesting that at least for some GRBs, the central engine has to be strongly magnetized (e.g. Zhang \& Pe'er 2009). 

The leading model of GRB central engine is a stellar-mass black hole (hereafter BH) surrounded by a neutrino-cooling-dominated accretion flow (hereafter NDAF) with an extremely high accretion rate (e.g. $0.01 - 1 M_\odot$/s). There are two main energy reservoirs to provide the jet power: the accretion energy in the disk that is carried by neutrinos and anti-neutrinos, which annihilate and power a bipolar outflow; and the spin energy of the black hole which can be tapped by a magnetic field connecting the outer world through the Blandford-Znajek (1977, hereafter BZ) mechanism. Both models have been extensively investigated by many authors (e.g., Popham et al. 1999; Lee et al. 2000; Li 2000; Narayan et al. 2001; Di Matteo et al. 2002; Kohri \& Mineshige 2002; Wang et al. 2002; McKinney 2005; Gu et al. 2006; Chen \& Beloborodov 2007; Janiuk et al. 2007; Lei et al. 2009). Some questions remain open: For example, which mechanism plays a more dominant role in jet power? Do they dominate in different luminosity regimes? Can one differentiate these different mechanisms using observational data?

In view of the recent observational constraints, we plan to systematically investigate the GRB BH central engine models in detail. The results are presented in two papers. In this first paper, we address a fundamental problem of GRB central engine: the baryon loading in the jet. All GRB prompt emission models rely on an assumed value of bulk Lorentz factor $\Gamma$. Recent broad band observations have led to measurement of $\Gamma$ for a good sample of GRBs, and interesting correlations between $\Gamma$ and the isotropic $\gamma$-ray energy and luminosity, i.e. $\Gamma \propto E_{\rm iso}^{\alpha_1} \propto L_{\rm iso}^{\alpha_2}$ with $\alpha_1 \sim \alpha_2 \sim (0.25-0.30)$, have been discovered (Liang et al. 2010; L\"u et al. 2012; cf. Ghirlanda et al. 2012). Most $\Gamma$ values were measured using the peak of the early afterglow light curve, which is believed to be related to the onset of the self-similar deceleration phase (e,g. Sari \& Piran 1999). This time is defined by the total energy in the ejecta and the density of the ambient medium, and essentially does not depend on the composition of the jet\footnote{L\"u et al. (2012) also discussed two other methods: the pair opacity constraint (Lithwick \& Sari 2001) and the upper limit of external shock emission during the prompt emission phase (Zou \& Piran 2010). These two methods are insensitive to the jet composition. The inclusion of these two methods do not lead to significant change of the slope of the $\Gamma - L_{\rm iso}$ correlation. }. The obtained correlation slope therefore does not pend on the jet power supply mechanisms. It is therefore interesting to investigate whether such a correlation roots from the fundamental physics of GRB baryon loading. This is the task of this paper. In the companion paper, we will investigate how the two jet mechanisms confront with the data of prompt GRB emission and early X-ray afterglow.

The paper is organized as follows. In Section 2, we calculate the neutrino-annihilation power and minimum baryon loading from a neutrino driven wind from a hyper-accreting BH disk. The role of magnetic field is ignored. By considering a range of BH spin, we simulate 2000 GRBs for their neutrino-annihilation jet energy (which is a proxy of the isotropic $\gamma$-ray energy) and the maximum Lorentz factor (defined by the minimum baryon loading). In Section 3, we consider a strongly magnetized BZ jet launched from the central BH.  We consider baryon loading into the magnetically dominated jet by the pickup neutron mechanism. Free neutrons can penetrate magnetic field lines and drift into the jet region. Through positron capture and proton-neutron collision avalanche, the neutrons are converted to protons and loaded in the jet. Our results are summarized in Section 4 with some discussion.

\section{Baryon Loading in a neutrino-annihilation powered jet}

The BH central engine model with a superaccreting disk has been studied extensively (e.g., Popham et al. 1999; Narayan et al. 2001; Di Matteo et al. 2002; Kohri \& Mineshige 2002; Gu et al. 2006; Chen \& Beloborodov 2007; Janiuk et al. 2007; Liu et al. 2007; Lei et al. 2009). In the inner region of such a hyperaccretion disk, a large amount of energetic neutrinos are emitted, which carry away the viscously dissipated energy of the accreted gas. If the accretion rate is large enough, cooling of the disk should be dominated by the neutrino emission, so that the disk is characterized as an  NDAF. 

For an NDAF with mass accretion rate $\dot{M}_{\rm ign}< \dot{M}<\dot{M}_{\rm trap}$, advection is not important. Here $\dot{M}_{\rm ign}$ and $\dot{M}_{\rm trap}$ are the critical accretion rates for igniting and suppressing neutrino cooling (Chen \& Beloborodov 2007). If $\dot{M} <\dot{M}_{\rm ign}$, the disc temperature is not high enough to ignite neutrino emitting reactions. If $\dot{M} > \dot{M}_{\rm trap}$, the emitted neutrinos become trapped in the disk and advected into the black hole. For the disk with viscosity $\alpha=0.1$, we find $\dot{M}_{\rm ign} = 0.071 M_{\sun} {\rm s^{-1}}$ and $\dot{M}_{\rm trap}=9.3 M_{\sun} {\rm s^{-1}}$ for $a_\bullet= 0$, and $\dot{M}_{\rm ign} = 0.021 M_{\sun} {\rm s^{-1}}$ and $\dot{M}_{\rm trap}=1.8 M_{\sun} {\rm s^{-1}}$ for $a_\bullet= 0.95$, where $a_{\bullet}=J_\bullet c/(G M_{\bullet}^2)$ is the spin parameter of a Kerr BH with mass $M_\bullet$ and angular momentum $J_\bullet$. In this case, the total neutrino power $\dot{E}_\nu$ can be approximated as
\begin{equation}
\dot{E}_{\nu} =\epsilon \dot{M}c^2 \simeq (1-E_{\rm ms}) \dot{M}c^2
\end{equation}
where $\epsilon$ is the neutrino emission efficiency, and $E_{\rm ms}$ is the specific energy corresponding to the inner edge radius $r_{\rm ms}$. The expression for $E_{\rm ms}$ is (Novikov \& Thorne 1973; Wang et al. 1998),
\begin{equation}
E_{\rm ms} = \frac{4\sqrt{ R_{\rm ms} }-3a_{\bullet}}{\sqrt{3} R_{\rm ms}},
\end{equation}
where $R_{\rm ms} = r_{\rm ms}/r_{\rm g}$ is the the radius of the marginally stable orbit in terms of $r_{\rm g} = G M_\bullet /c^2$. We have $0.06< \epsilon <0.42$ for $0< a_\bullet <1$. The radius $R_{\rm ms}$ is expressed as (Bardeen et al. 1972; Page \& Thorne 1974),
\begin{eqnarray}
R_{\rm ms} =  3+Z_2 -\left[(3-Z_1)(3+Z_1+2Z_2)\right]^{1/2},
\end{eqnarray}
for $0\leq a_{\bullet} \leq 1$, where $Z_1 \equiv 1+(1-a_{\bullet}^2)^{1/3} [(1+a_{\bullet})^{1/3}+(1-a_{\bullet})^{1/3}]$, $Z_2\equiv (3a_{\bullet}^2+Z_1^2)^{1/2}$. We have $R_{\rm ms} =6.0$ for $a_\bullet=0$, and $R_{\rm ms} =2.3$ for $a_\bullet=0.9$. 

The neutrino annihilation ($\nu \bar{\nu} \rightarrow e^{+} e^{-}$) process can launch a relativistic jet reaching the GRB luminosity. For a system with black hole mass $M_\bullet$ and spin $a_\bullet$, the neutrino annihilation power $\dot{E}_{\nu\bar{\nu}}$ from the NDAF depends on the accretion rate $\dot{M}$.  For $\dot{M}_{\rm ign}< \dot{M}<\dot{M}_{\rm trap}$, the neutrino annihilation power can be approximated as (Zalamea \& Beloborodov 2011),
\begin{equation}
\dot{E}_{\nu \bar{\nu}} \simeq 1.1 \times 10^{52} \left(\frac{R_{\rm ms} }{2} \right)^{-4.8} \left(\frac{m}{3}\right)^{-3/2} \dot{m}^{9/4} ~{\rm erg \ s^{-1}},
\label{eq:Evv}
\end{equation}
where $m=M_{\bullet}/M_{\sun}$, and $\dot{m} = \dot{M}/M_{\sun} \rm s^{-1}$. 

Neutrino heating in the atmosphere just above the disk surface results in mass-loss from the hyperaccreting disk. The dominant heating processes are electron neutrino absorption on baryons ($p+\bar{\nu}_e \rightarrow n+e^+$ and $n+\nu_e \rightarrow p+e^-$). For an unmagnetized neutrino-driven wind, the mass-loss rate $\dot{M}_{\nu}$ can be estimated as (see also Metzger et al. 2008; Qian \& Woosley 1996)
\begin{equation}
\dot{M}_{\nu} \simeq 10^{-6}  \dot{E}_{\nu,52}^{5/3} \langle\epsilon_{10}^2\rangle^{5/3} r_6^{5/3} (m/3)^{-2}(h/r)^{-1} M_{\sun} {\rm s}^{-1}.
\label{eq:Mv}
\end{equation}
where $h$ is the half-thickness of disk, $r$ is the disk radius, $\epsilon_{\nu} = \epsilon_{10} \times 10 {\rm MeV}$ is the energy of neutrinos, $\langle\epsilon_\nu^2\rangle = 13.8 (k T)^2$, and $T$ is the disk temperature (Di Matteo et al. 2002). Hereafter, the convention $Q_n = Q/10^n$ in cgs unit is adopted if not otherwise defined.

The disk temperature $T$ and height $h$ can be obtained by solving the set of equations describing NDAF. Since we are interested in the inner disk with moderate high accretion rate $\dot{M}_{\rm ign}< \dot{M}<\dot{M}_{\rm trap} $, the disk is quite dense and hot. As a result, cooling by pair capture on nucleons $q^-_{\rm eN}$ should dominate over $\nu-\bar{\nu}$ annihilation. The energy and angular momentum equations as well as the equation of state can be simplified as (e.g. Reynoso, Romero {\&} Sampayo 2006; Lei et al. 2009; Liu et al. 2010):
\begin{equation}
\label{E_Kerr}
\frac{3GM_{\bullet} \dot {M}}{8\pi r^3}\frac{D}{B} \simeq q_{\rm eN}^- h =9.0\times 10^{-43} \rho T^6 X_{\rm{nuc}} h
\end{equation}

\begin{equation}
\dot {M}r^2\sqrt {\frac{GM_\bullet }{r^3}} \frac{D}{A} = 4\pi r^2 h\alpha P\sqrt {\frac{A}{BC}}
\end{equation}

\begin{equation}
P = P_{\rm gas} + P_{\rm rad} + P_{\rm deg} + P_{\rm \nu} \simeq P_{\rm gas} = \frac{\rho kT}{m_{\rm{p}} }(\frac{1 + 3X_{\rm{nuc}} }{4})
\end{equation}
where $\rho$ is the disk density. The total pressure $P$ consists of four terms: gas pressure $P_{\rm gas}$, radiation pressure $P_{\rm rad}$,  degeneracy pressure $P_{\rm deg}$, and neutrino pressure $P_{\rm \nu}$. In the inner disk region, one generally has $P_{\rm gas}$ dominating other terms (e.g. Di Matteo et al. 2002). The parameter $X_{\rm{nuc}} $ is the mass fraction of free nucleons. In the inner disk, it is found $X_{\rm{nuc}} \simeq 1$ (Popham et al. 1999). Hydrostatic equilibrium in the vertical direction leads to a corrected expression for the half thickness of the disk (Riffert {\&} Herold 1995; Reynoso, Romero {\&} Sampayo 2006).
\begin{equation}
h \simeq \sqrt { \frac{ Pr^3} {\rho GM}  \frac{B}{C} }.
\label{eq:h}
\end{equation}

In Equations (\ref{E_Kerr})-(\ref{eq:h}), the relativistic correction factors for a thin accretion disk around a Kerr BH are given by Riffert {\&} Herold (1995), i.e.
\begin{eqnarray}
\label{KerrFactors}
&& A = 1 - 2R^{-1} + a_\bullet^2 R^{-2}, \  B = 1 - 3R^{-1} + 2a_\bullet R^{-3 / 2}, \nonumber \\
&& C = 1 - 4a_\bullet R^{-3 / 2} + 3 a_\bullet^2 R^{-2}, \nonumber \\
&& D = \int_{R_{\rm{ms}} }^R {\frac{ x^2 - 6 x + 8 a_\bullet x^{1/2} - 3 a_\bullet ^2 }{  2 \sqrt{Rx} (x^2 - 3x + 2a_\bullet x^{1/2} )}dx} .
\end{eqnarray}
where $R=r/r_{\rm g}$ is the disk radius in terms of $r_{\rm g}$.

Combining Equations (\ref{E_Kerr}) - (\ref{KerrFactors}), we get
\begin{equation}
T \simeq 1.2 \times 10^{11} A^{0.3} B^{-0.3} C^{-0.1} \alpha^{0.2} m^{-0.2} R^{-0.3} \rm K,
\end{equation}
\begin{equation}
h \simeq 1.6 \times 10^4 A^{0.15} B^{0.35} C^{-0.55} \alpha^{0.1} m^{0.9} R^{1.35} \rm cm.
\end{equation}

A neutrino-annihilation-powered jet has an opening angle of $\theta_{\nu\bar{\nu}} \simeq 0.1$ (Aloy, Janka \& Muller 2005; Harikae et al. 2010). Considering only the neutrino wind that enters this funnel, the baryon loading rate of the jet can be estimated as
\begin{eqnarray}
\dot{M}_{\rm j, \nu\bar{\nu}} & = &  \dot{M}_{\nu} \theta_{\nu\bar{\nu}}^2/2 \nonumber \\ 
& = & 7.0 \times 10^{-7} A^{0.85} B^{-1.35} C^{0.22} \theta_{\nu\bar{\nu},-1}^2  \alpha_{-1}^{0.57} \epsilon_{-1}^{1.7} \nonumber \\
& & \left( \frac{R_{\rm ms} }{2} \right) ^{0.32} \dot{m}_{-1}^{1.7} \left(\frac{m}{3}\right)^{-0.9} \left(\frac{\xi}{2} \right)^{0.32} \ M_{\sun} {\rm s}^{-1}.
\end{eqnarray}
where $\xi \equiv r/r_{\rm ms}$ is the disk radius in terms of $r_{\rm ms}$. 

Let us define a dimensionless ``entropy'' parameter
\begin{equation}
\eta \equiv \frac{\dot{E}_{\rm m}}{ \dot{M}_{\rm j, \nu\bar{\nu}} c^2 }.
\label{eta}
\end{equation}
where $\dot{E}_{\rm m} = \dot{E}_{\nu \bar{\nu}}+ \dot{M}_{\rm j, \nu\bar{\nu}} c^2$ is the total matter energy outflow luminosity. If most neutrino annihilation energy is converted into kinetic energy of baryons after acceleration, and the jet would reach a Lorentz factor $\Gamma_{\rm max} \simeq \eta$. 

The value of the parameter $\eta$ likely changes during a GRB, since the BH has spin evolution during the hyperaccretion process. Without magnetic fields, the dominant mechanism is spin-up due to accretion. The process can be delineated by
\begin{equation}
\frac{dM_\bullet c^2}{dt} = \dot{M} c^2 E_{\rm ms},
\label{eq:dMvv}
\end{equation}
\begin{equation}
\frac{dJ_\bullet}{dt} = \dot{M} L_{\rm ms},
\label{eq:dJvv}
\end{equation}
where $L_{\rm ms}$ are the specific angular momentum corresponding to the inner most radius $r_{\rm ms}$ of the disk, which is defined as (Novikov \& Thorne 1973) 
\begin{equation}
L_{\rm ms} = \frac{G M_\bullet}{c} \frac{2 (3 \sqrt{R_{\rm ms}} -2 a_\bullet) }{\sqrt{3} \sqrt{R_{\rm ms}} }.
\end{equation}
Since $a_\bullet = J_\bullet c/(GM_\bullet^2)$, by incorporating the above two Equations (\ref{eq:dMvv}) and (\ref{eq:dJvv}), the evolution of the BH spin can be expressed by
\begin{equation}
\frac{da_\bullet}{dt} =  \dot{M} L_{\rm ms} c/(GM_\bullet^2) -2 a_\bullet \dot{M} c^2 E_{\rm ms} /(M_\bullet c^2).
\end{equation}
Considering spin evolution, one can define an average $\eta$ during the evolution of a GRB, i.e.
\begin{equation}
\bar\eta =  \frac{\int \dot{E}_{\rm m} dt}{\int \dot{M}_{\rm j, \nu\bar{\nu}} c^2 dt}.
\end{equation}

For a hot fireball, the $\eta$ parameter is related to the terminating Lorentz factor if $\eta$ is not too high (e.g. M\'esz\'aros \& Rees 2000). Recent observations of GRBs have led to constraints of GRB Lorentz factor for a sample of GRBs (e.g. Liang et al. 2010; L\"u et al. 2012 and references therein). By constraining $\Gamma$ of about 20 GRBs through modeling the deceleration bump feature in the early afterglow lightcurves, Liang et al. (2010) discovered a tight correlation between $\Gamma$ and $E_{\rm \gamma,iso}$, i.e. $\Gamma \simeq 182 (E_{\rm \gamma,iso}/10^{52}{\rm erg})^{0.25}$. L\"u et al. (2012) confirmed the $\Gamma-E_{\rm \gamma,iso}$ correlation (Liang et al. 2010) with an extended sample (about 50 GRBs) by applying more methods to constrain $\Gamma$. They also discovered an even tighter correlation $\Gamma \simeq 249 L_{\rm \gamma,iso,52}^{0.30}$, where $L_{\rm \gamma,iso}$ is the mean luminosity of the burst. In L\"u et al. (2012), we have proposed that $\Gamma \propto L_{\rm \gamma,iso}^{0.30}$ can be explained within the BH-NDAF GRB central engine model using a simplified model. Here we give much more detailed modeling by including the effect of BH spin.

\begin{figure}[htc]
\center
\includegraphics[width=7cm,angle=0]{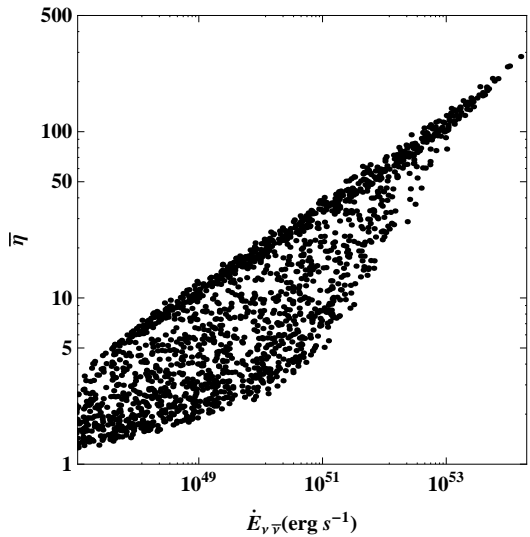}
\caption{$\bar\eta$ vs. the central engine output power $\dot{E}_{vv}$. Two thousand GRBs with random values of BH spin, BH mass, accretion rate and disk mass are simulated. The best fit gives $\bar\eta \propto \dot{E}_{\nu\bar{\nu}}^{0.27}$.}
\label{fig1}
\end{figure}

In Fig.\ref{fig1}, we show the simulated 2000 GRBs with random values of BH spin $a_\bullet$, BH mass $m$, accretion rate $\dot{m}$ and disk mass $m_{\rm d}$. Other parameters take the typical values. We allow $a_\bullet$, $m$, $\dot{m}$ and $m_{\rm d}$ to randomly vary in the range of (0.1, 0.998), (3, 10), (0.01, 3) and (0.1, 30) respectively. We adopt a logarithmic distribution for the accretion rate, while linear distributions for other parameters.

By fitting Fig.\ref{fig1}, we find a best-fit correlation $\bar\eta \propto \dot{E}_{\nu\bar{\nu}}^{0.27}$. Since $\eta$ values are typically lower than several 100s, it is reasonable to believe that $\bar\eta$ is essentially the bulk Lorentz factor $\Gamma$, so that $\Gamma \simeq \bar\eta \propto \dot{E}_{\nu\bar{\nu}}^{0.27}$. This result is nearly the same as what we got in L\"u et al. (2012), in which the index is $7/27$. 

In order to compare with the observations, we need to consider two effects. One is the $\gamma$-ray radiation efficiency $\eta_\gamma$. The other is the beaming effect. For simplicity, we assume a relatively constant $\eta_\gamma$, so that $L_{\gamma} = \eta_{\gamma} \dot{E}_{\nu \bar{\nu}} \propto \dot{E}_{\nu \bar{\nu}}$, and $\Gamma \propto L_{\rm \gamma}^{0.27}$. The isotropic luminosity $L_{\rm \gamma,iso}$ and $L_\gamma$ are connected through the beaming factor $f_b \ll 1$, i.e. $L_\gamma = f_b L_{\rm \gamma,iso}$.  By combining the Amati relation $E_p^{\prime} \propto E_{\gamma,{\rm iso}}^{0.57}$ (Amati et al. 2002, 2008; Amati 2006) and the Girlanda relation $E_{\gamma} \propto (E_p^{\prime} )^{3/2} $ (Ghirlanda et al. 2004), one may obtain a relation between $f_b$ and $E_{\gamma,{\rm iso}}$, i.e. $f_b \propto E_{\gamma,{\rm iso}}^{-0.145}$. Since $L_{\gamma,{\rm iso}} \propto E_{\gamma,{\rm iso}}$, one can get $f_b \propto L_{\gamma,{\rm iso}}^{-0.145}$, which is very insensitive to $L_{\rm \gamma,iso}$ and $E_{\rm \gamma,iso}$. We then obtain the relation between Lorentz factor $\Gamma$ and the isotropic luminosity $L_{\rm \gamma,iso}$ 
\begin{eqnarray}
\Gamma \propto \dot{E}^{0.27} \propto (f_b L_{\rm \gamma,iso})^{0.27} \propto L_{\rm \gamma,iso}^{0.23}~,
\end{eqnarray}
which agrees well with the statistical correlations obtained by Liang et al. (2010) ($\Gamma \propto (E_{\rm \gamma,iso})^{0.25}$) and L\"u et al. (2012) ($\Gamma \propto  L_{\rm \gamma, iso}^{0.30}$). We note that the insensitive $f_b$ on $L_{\rm \gamma,iso}$ is crucial to the above argument.

We'd like to caution that despite of the correct power law index of the correlation, the absolute values of $\eta$ (and hence, $\Gamma$) is typically a little bit too low. Inspecting Fig.\ref{fig1}, one can see that for $\dot E_{\nu\bar\nu} = 10^{51}~{\rm erg~s^{-1}}$, the Lorentz factor $\Gamma$ is $\leq 40$. In order to reach large $\Gamma$ values, one needs to appeal to model parameters that invoke large spin and high accretion rate. Another possibility to reconcile with the data is to assume a relatively small $\eta_\gamma$, so that for an observed $L_{\gamma,iso}$, the corresponding $\dot E_{\nu\bar\nu}$ is much larger, which corresponds to a higher $\Gamma$ to be consistent with the data.

\section{Baryon loading in a Blandford-Znajek jet}

With magnetic field lines threading the horizon of a Kerr black hole, the rotational energy of the black hole can be extracted by the BZ mechanism (Blandford \& Znajek 1977). The BZ jet power from a BH with mass $M_{\bullet}$ and angular momentum $J_\bullet$ is (Lee et al. 2000; Li 2000; Wang et al. 2002; McKinney 2005; Lei \& Zhang 2011) 
\begin{equation}
\dot{E}_{\rm B}=1.7 \times 10^{50} a_{\bullet}^2 m^2 
B_{\bullet,15}^2 F(a_{\bullet}) \ {\rm erg \ s^{-1}},
\label{eq_Lmag}
\end{equation}
where $B_{\bullet,15}=B_{\bullet}/10^{15} {\rm G}$ and
\begin{equation}
F(a_{\bullet})=[(1+q^2)/q^2][(q+1/q) \arctan q-1]
\label{eq_F}
\end{equation}
here $q= a_{\bullet} /(1+\sqrt{1-a^2_{\bullet}})$, and $2/3\leq F(a_{\bullet}) \leq \pi-2$ for 
$0\leq a_{\bullet} \leq 1$. It apparently depends on $M_{\bullet}$, $B_{\bullet}$,
and $a_{\bullet}$. A strong magnetic field of $\sim 10^{15} \rm G$ is required to produce the high luminosity of a GRB.

As the magnetic field on the BH is supported by the surrounding disk, there are some relations between $B_{\bullet}$ and $\dot{M}$. In a hyper-accreting flow in a GRB, it is possible that a magnetic flux is accumulated near the black hole horizon. Considering the balance between the magnetic pressure on the horizon and the ram pressure of the innermost part of the accretion flow (e.g. Moderski et al. 1997), one can estimate the magnetic field strength threading the BH horizon
\begin{equation}
\frac{B_{\bullet}^2}{8\pi} = P_{\rm ram} \sim \rho c^2 \sim \frac{\dot{M} c}{4\pi r_{\bullet}^2}
\label{Bmdot}
\end{equation}
where $r_{\bullet}=(1+\sqrt{1-a_\bullet^2})r_{\rm g}$ is the radius of the BH horizon. It can be rewritten as 
\begin{equation}
B_{\bullet} \simeq 7.4 \times 10^{16} \dot{m}^{1/2} m^{-1} (1+\sqrt{1-a_\bullet^2})^{-1} \rm{G}.
\end{equation} 
Inserting it to Equation (\ref{eq_Lmag}), we obtain the magnetic power as a function of mass accretion rate and BH spin, i.e.
\begin{equation}
\dot{E}_{\rm B}=9.3 \times 10^{53} a_\bullet^2 \dot{m}  X(a_\bullet) \ {\rm erg \ s^{-1}} ,
\end{equation}
and
\begin{equation}
X(a_\bullet)=F(a_\bullet)/(1+\sqrt{1-a_\bullet^2})^2.
\end{equation}
It is found that $X(0)=1/6$, and $X(1)=\pi -2$. In general, a faster BH spin is more favorable for GRB production, as revealed also by recent GRMHD numerical simulations (Nagataki 2009, 2011).

We next consider baryon loading in a BZ jet. The BH magnetosphere has a ``floor'' charge density defined by the force-free condition (Goldrich \& Julian 1969). We use it to define a minimum baryon loading rate of a BZ jet. It is given by
\begin{eqnarray}
\dot{M}_{\rm GJ} &=&  7\times 10^{-2} (m_p+\varsigma m_e) 4\pi r_\bullet^2 B_\bullet \Omega_{\rm F} c/2\pi \nonumber \\ 
&=& 2.9\times 10^{-16} (1+0.5\varsigma_3) a_\bullet \dot{m}^{1/2} M_\sun s^{-1},
\end{eqnarray}
where $\varsigma_3=\varsigma/10^3$, and $\varsigma$ is multiplicity of electron-positron pairs, which is rather uncertain. Here $\Omega_{\rm F}=0.5\Omega_\bullet$ is usually taken to maximize the BZ power, and
\begin{equation}
\Omega_\bullet = \frac{c^3}{G M_\bullet} \frac{a_\bullet}{2 (1+\sqrt{1-a_\bullet^2})}
\end{equation}
is the angular velocity of the BH horizon.

Since the magnetic field is supported by the accretion disk, in the BZ model, the BH must be surrounded by a hyperaccreting NDAF. The physical processes discussed in the previous section must still happen, which tend to load baryons into the jet. The main difference is that the magnetic field threading the BH makes a strong magnetic barrier that prevents charged baryons (protons) to enter the jet, making a baryon poor jet (e.g. Li 2000). This baryon-poor jet is surrounded and collimated by an optically thick baryonic outflow from the hyperaccreting disk (Eichler \& Levinson 1999; Levinson \& Eichler 2003). Hereafter we assume that the dominant source for the baryons is the neutrino-driven wind from the hyperaccreting disk. Other baryon loading processes may happen, e.g. baryon contamination from the sideways by instabilities during the propagation of the jet, or baryons entrained from the magnetic loops erupted from the disk that may enter the jet region (e.g. Yuan \& Zhang 2012). So our model gives the \textit{minimum} baryon loading in a magnetized BZ jet.

A strong magnetic field may change $\dot{M}_\nu$ by altering the neutrino heating and cooling rates in the hyperaccreting disk (Zhang \& Dai 2010). The most important effect is that electrons and positrons participating in the charged-particle reactions are restricted into discrete Landau levels (Duan \& Qian 2004). For first order estimation, in this paper we neglect these effects.

Since the magnetic field only affects the charged particles, it is important to first study the composition of the wind. According to Pruet, Woosley \& Hoffman (2003) and Chen \& Beloborodov (2007), a hyperaccretion flow is neutron-rich in its inner region. The fraction of protons $f_{\rm p}$ is only around $0.1$, so that the majority of baryons are neutrons (Chen \& Beloborodov 2007). The neutrino-driven wind would take the similar mass composition. The number density of neutrons in the wind can be therefore expressed as
\begin{eqnarray}
n_{\rm n} & \simeq & \frac{\dot{M}_{\nu} f_{\rm n}}{4\pi r_z^2 v_{\rm w} m_{\rm p} } \nonumber \\
& \simeq & 3.5 \times 10^{20} A^{0.85} B^{-1.35} C^{0.22} f_{\rm n} \alpha_{-1}^{0.57} \nonumber \epsilon_{-1}^{1.7} R_{\rm ms}^{0.32} \\
& &  \dot{m}_{-1}^{1.7} \left(\frac{m}{3}\right)^{-0.9} \left(\frac{\xi}{2} \right)^{0.32} r_{z,11}^{-2} \beta_{\rm w,-1}^{-1} \ \rm cm^{-3},
\label{eq:nn}
\end{eqnarray} 
where $f_{\rm n}=1-f_{\rm p}$ is the fraction of neutrons, $\epsilon \equiv 0.1 \epsilon_{-1}$ is neutrino emission efficiency (Eq.(1)), $r_z$ is the distance from the BH in the jet direction, which is normalized to $10^{11}$ cm, the typical radius of the progenitor star, $\beta_{\rm w, -1}=\beta_{\rm w}/0.1$, and $v_{\rm w} = \beta_{\rm w} c$ is the wind speed. These neutrons can penetrate magnetic field lines and freely fill any location above the disk.

Protons are different. Because of the existence of magnetic fields, only the neutrino-driven outflow in preferred directions, i.e. almost align with the magnetic field lines from the field line foot on the disk, can be ejected into the atmosphere. Those protons with an ejected direction larger than an angle $\theta_{\rm B}$ with respect to the field lines would be blocked. For a rough estimate, the proton density in the region where a local field line connects with the disk (which is relevant for regions outside the BZ jet) can be estimated as

\begin{eqnarray}
n_{\rm p} & \simeq & \frac{\dot{M}_{\nu} f_{\rm p} \theta_{\rm B}^2 }{4\pi r_z^2 v_{\rm w} m_{\rm p} } \nonumber \\
& \simeq & 3.5 \times 10^{15} A^{0.85} B^{-1.35} C^{0.22} f_{\rm p,-1} \theta_{\rm B,-2}^2 \alpha_{-1}^{0.57} \epsilon_{-1}^{1.7} \nonumber \\
& &  R_{\rm ms}^{0.32} \dot{m}_{-1}^{1.7} \left(\frac{m}{3}\right)^{-0.9} \left(\frac{\xi}{2} \right)^{0.32} r_{z,11}^{-2} \beta_{\rm w, -1}^{-1} \ \rm cm^{-3}.
\label{eq:np}
\end{eqnarray} 

Free protons and neutrons in the wind are coupled by nuclear elastic scattering. At the temperatures of interest, the corresponding rate is $\langle \sigma_{el} v \rangle \simeq 10^{-15} \rm cm^3s^{-1}$, independent of the center-of-mass energy. For a neutron, the optical depth for elastic scattering with protons is
\begin{eqnarray}
\tau_{np} & \simeq  & n_p \sigma_{el} r_z \nonumber \\
& \simeq & 172 A^{0.7} B^{-1.2} C^{0.27} f_{\rm p,-1} \theta_{\rm B,-2}^2 \alpha_{-1}^{0.47} \nonumber \\
& & \epsilon_{-1}^{1.7} R_{\rm ms}^{0.47} \dot{m}_{-1}^{1.7} \left(\frac{m}{3}\right)^{-0.8} \left(\frac{\xi}{2} \right)^{0.47} r_{z,11}^{-1} \beta_{\rm w, -1}^{-1},
\end{eqnarray}

During the propagation of the jet, neutrons drift from sideways into the jet. The flux of neutrons diffusing into the magnetized jet is $J_{\rm D}(r) = \lambda_{\rm np} v_{\rm n} \partial n_{\rm n} /\partial x = \lambda_{\rm np} v_{\rm n} (n_{\rm n} /l)$, where $\lambda_{\rm np}=1/(n_{\rm p} \sigma_{\rm el})$ is the mean free path of $n-p$ collisions, $x$ denotes the cylindrical radius, $l\simeq (\lambda_{\rm np} v_{\rm n} t_{\rm exp})^{1/2}$ denotes the gradient length scale, $t_{\rm exp}=r/v_{\rm n}$ is the wind expansion time, and $v_{\rm n} = \beta_{\rm n} c = (kT/m_{\rm p})^{1/2}$ is the neutron thermal speed. For $T=10^{11}$K, we have $v_{\rm n} \sim 0.1c$. For a typical jet opening angle $\theta_{\rm BZ} \sim 0.1$,  the neutron drift rate into the jet is

\begin{eqnarray}
\dot{M}_{\rm n} & = & 2\pi \theta_{\rm BZ} r_z^2 J_{\rm D}(r) \nonumber \\
& \simeq & 3.5 \times 10^{-7} A^{0.58} B^{-0.83} f_{\rm p,-1}^{-0.5} \theta_{\rm BZ,-1} \theta_{\rm B,-2}^{-1} \nonumber \\
& &  \alpha_{-1}^{0.38} \epsilon_{-1}^{0.83} \dot{m}_{-1}^{0.83} \left(\frac{m}{3}\right)^{-0.55}  r_{z,11}^{0.5} \ M_{\sun} {\rm s}^{-1}.
\end{eqnarray}

The neutrons that enter the jet are not accelerated magnetically. In order to be loaded in the jet, neutrons should be converted to protons. The first mechanism would be free neutron decay. The rest-frame decay time scale is $t_{\rm decay} \sim 900 \rm s$, which corresponds to a typical decay radius of $\sim 10^{15}$ cm. This is not an effective mechanism to load baryons near the central engine. Below we consider following two mechanisms that can quickly convert a significant fraction of neutrons to protons.

The first mechanism is positron capture. As shown in Section 2, a hyperaccreting disk produces a strong neutrino/anti-neutrino wind that deposit electron-positron pairs in the magnetized jet via neutrino annihilation ($\nu \bar{\nu} \rightarrow e^{+} e^{-}$). This leads to proton production via
\begin{equation}
e^{+} + n \rightarrow p + \bar{\nu}_e.
\end{equation}
Dividing the neutrino annihilation power $\dot{E}_{\nu\bar{\nu}}$ (Equation (\ref{eq:Evv})) by the average neutrino energy $\langle \epsilon_{\nu} \rangle \sim 10$ MeV, we obtain a rough estimate of the number rate of $e^{+}e^{-}$ pairs: $\dot{N}_{e^{+}e^{-}} \sim 7 \times 10^{56} \rm s^{-1}$. The number density of pairs is around $n_{e^{+}e^{-}} \simeq   \dot{N}_{e^{+}e^{-}}/ (\pi r_z^2 c) \sim 7.3 \times 10^{31} \rm cm^{-3}$.

The rate of $e^+$ capture can be derived from the standard electroweak theory (e.g., Shapiro \& Teukolsky 1983; Bruenn 1985). In the jet region, the number density of nucleons are low enough to satisfy the non-degenerate condition. One then obtains the positron capture rate
\begin{equation}
\dot{n}_{e^+ n} =  K n_n \int_{Q+1}^{\infty} f_+ (\omega - Q) (\omega - Q)^2 \sqrt{1-\frac{1}{(\omega - Q)^2}} \omega^2 d\omega,
\label{dne}
\end{equation}
where $\omega$ is neutrino energy in units of $m_e c^2$, $Q=(m_n-m_p)/m_e = 2.531$, and $K\simeq 6.5 \times 10^{-4} s^{-1}$. The function $f_+(\omega-Q)$ is the Fermi-Dirac distribution
\begin{equation}
f_+(\omega-Q) = \frac{1}{\exp[((\omega-Q)-\mu_+)/\theta_e] + 1},
\end{equation}
where $\theta_e = k T_e/m_e c^2$ and $\mu_+$ is the positron chemical potential in units of $m_e c^2$.

At $\mu_+ < \theta_e$ and $\theta_e > Q+1$, equation (\ref{dne}) is simplified as
\begin{equation}
\dot{n}_{e^+ n} = K n_n \theta_e^5 \left[\frac{45}{2} \zeta(5) - \frac{7 \pi^4}{60} \frac{(2\mu-Q)}{\theta_e} \right],
\end{equation}
where $\zeta(5)=1.037$ is the Riemann $\zeta$-function. Here, we neglected the next-order terms $O(Q^2/\theta_e^2)$, $O(\mu^2/\theta_e^2)$, and $O[(Q+1)^5/\theta_e^5]$ and used the formula $\int_0^\infty (\exp(x)+1)^{-1}x^n dx = (1-2^n)\times \Gamma(n+1)\zeta(n+1)$ with $\Gamma(n+1) = n!$ for integer $n$. The above equation can be further simplified (in zero order in $\mu/\theta_e$) as 
\begin{equation}
\dot{n}_{e^+ n} \simeq 1.5\times 10^{-2} n_n \theta_e^5 \rm cm^{-3} s^{-1}.
\end{equation}
The temperature of electrons produced by neutrino-antineutrino annihilation is given by
\begin{equation}
\frac{1}{3} a T_e^4 \pi r_z^2 \beta_{j} c = \dot{E}_{\nu\bar{\nu}},
\end{equation}
where $\beta_j$ is the jet velocity.

The timescale of neutron capture can be estimated by
\begin{eqnarray}
t_{\rm cap} \simeq \frac{n_n}{\dot{n}_{e^+ n}} = 7 \left(\frac{R_{\rm ms}}{2}\right)^6 r_{z,7}^{2.5} \beta_{j,-1}^{1.25} \dot{m}_{-1}^{-2.8} \ \rm s.
\label{eq:tcap} 
\end{eqnarray}
The capture time is sensitive to the distance $r_z$ from the BH. Close to the BH, a large fraction of neutrons that drifted into the BZ jet would be captured. This fractions drops quickly with increasing $r_z$, since the density and temperature of electrons are lower at larger distances. As a result, we normalize $r_z$ to $10^7$ cm, the typical size of the central engine.

The second, maybe more efficient mechanism to convert neutrons to protons is proton-neutron \textit{inelastic} collision avalanche (Levinson \& Eichler 2003). Protons entrained in the jet would be accelerated magnetically and soon reach an energy large enough so that inelastic collisions with neutrons would happen. These collisions ($pn \rightarrow pp\pi^+ ...$, $pn \rightarrow pp\pi^- $) efficiently convert neutrons to protons (and other way round) so that the proton and neutron fractions become comparable. A proton produced through positron capture would generate more protons via inelastic collisions with the neutrons. The proton fraction thus grows exponentially in what we term as a collision avalanche, until a proton-neutron equilibrium is reached. The optical depth for an inelastic collision of a picked-up proton with the target neutrons is 
\begin{eqnarray}
\tau_{n-p} & \simeq & \sigma_{n-p} r_z \dot{M}_{\rm n} /(\pi \theta_{\rm BZ}^2 r_z^2 v_j m_p) \nonumber \\
& \simeq & 1.8 \times 10^6 A^{0.58} B^{0.83} f_{\rm p,-1}^{-0.5} \theta_{\rm BZ,-2}^{-1}  \nonumber \\
& & \theta_{\rm B,-2}^{-1}  \alpha_{-1}^{0.38} \epsilon_{-1}^{0.83} \dot{m}_{-1}^{0.83} (\frac{m}{3})^{-0.55}  r_{z,11}^{-0.5} \beta_j^{-1}.
\end{eqnarray}
where $\sigma_{n-p}=40 ~{\rm mbarn}$ for inelastic collision (Hagiwara et al. 2002). 

The remaining free neutrons would decouple from protons at a larger radius (Derishev et al. 1999; M\'esz\'aros \& Rees 2000). The free neutrons would decay at a larger radius (e.g. $R \sim 10^{15}$ cm) and eventually picked up by the jet, and leaves some observational signatures in the early afterglow phase (Beloborodov 2003; Fan et al. 2005). Eventually, all the neutrons drifted into the jet are loaded in the jet. In the following discussion, we estimate baryon loading rate in a BZ jet as the neutron drifting rate into the jet, i.e. $\dot{M}_{\rm j, BZ} \simeq \dot{M}_{\rm n}$.

For a magnetized central engine, one can define a parameter
\begin{equation}
\mu_0 \equiv \frac{\dot{E}}{\dot{M}_{\rm j,BZ} c^2} = \frac{\dot{E}_{\rm m}+\dot{E}_{\rm B}}{\dot{M}_{\rm j,BZ} c^2} = \eta (1+\sigma_0),
\label{eq:mu}
\end{equation}
where $\dot{E}_{\rm m} = \dot{E}_{\nu \bar{\nu}}+ \dot{M}_{\rm j,BZ} c^2$, and $\sigma_0 = \dot E_{\rm B}/\dot E_{\rm m}$. This parameter denotes the maximum available energy per baryon in the jet.

The detailed acceleration process of a jet with both thermal power and magnetic power has not been studied in detail. In general, thermal acceleration proceeds much faster than magnetic acceleration, so that the initial matter power $\dot E_{\rm m}$ would be quickly converted to a matter flux, so that $\sigma_0$ carries the usual definition of the ratio between a Poynting flux and a matter flux. If no magnetic dissipation occurs, the parameter 
\begin{equation}
 \mu = \mu_0 = \eta (1+\sigma_0) = \Gamma (1+\sigma)
\label{mu}
\end{equation}
remains a constant as magnetic acceleration proceeds. $\Gamma$ continues to increase while $\sigma$ drops (e.g. Komissarov et al. 2009; Tchekhovskoy et al. 2010). An efficient magnetic acceleration terminates when a causal contact condition is broken. This occurs as the bulk flow Lorentz factor reaches the Alfven Lorentz factor $\Gamma_{\rm A} = (1+\sigma)^{1/2} \sim \sigma^{1/2}$. According to Equation (\ref{mu}), this corresponds to $\sigma \sim \mu_0^{2/3}$ and $\Gamma \sim \mu_0^{1/3}$. Combining thermal acceleration and efficient magnetic acceleration, the outflow would quickly reach a Lorentz factor

\begin{equation}
\Gamma_0=\max(\mu_{0}^{1/3},\eta)
\label{Gamma0}
\end{equation}
at a radius $r_0 \simeq 2\times 10^{11} \rm cm$, which is not far beyond the radius of progenitor's envelope.

The acceleration behavior of the jet beyond $r_0$ is subject to uncertainties. For a non-dissipative, steady magnetized outflow, the acceleration is rather inefficient (e.g. Lyubarsky 2010). The most efficient acceleration may be proceeded as 
\begin{equation}
\Gamma \sim \Gamma_0 (r/r_0)^{1/3}, \ \ \sigma \sim \sigma_0 (r/r_0)^{-1/3}
\label{eq_gammar}
\end{equation}
if there is a continuous magnetic dissipation due to current instability (Giannios \& Spruit 2006) or the pulse of the shell is short enough to undergo an ``impulsive'' acceleration (Granot et al. 2011). Depends on the initial $\mu_0$, the jet may or may not reach the full Lorentz factor 
\begin{equation}
\Gamma_{\rm max} = \mu_0.
\end{equation}
Rather, more likely the unsteady jet (as manifested by the erratic lightcurve behavior) would undergo internal collisions, which lead to distortion of magnetic configurations and trigger an Internal-Collision induced MAgnetic Reconnection and Turbulence (ICMART) avalanche to discharge the magnetic energy (Zhang \& Yan 2011). Such a process would make the outflow to reach a terminating Lorentz factor $\Gamma$ that satisfies
\begin{equation}
 \Gamma_0 < \Gamma < \Gamma_{\rm max},
\label{gamma}
\end{equation}
with the explicit value depending on the detailed dissipation process.

The central engine parameters evolve with time during a GRB, since the BH would be spun-up by accretion while spun-down by the BZ mechanism. The evolution equations of a Kerr BH in the BZ model can be written as
\begin{equation}
\frac{dM_\bullet c^2}{dt} = \dot{M} c^2 E_{\rm ms} - \dot{E}_{\rm B},
\label{dMbz}
\end{equation}

\begin{equation}
\frac{dJ_\bullet}{dt} = \dot{M} L_{\rm ms} - T_{\rm B}.
\label{dJbz}
\end{equation}
The evolution equation of the BH spin is then
\begin{eqnarray}
\frac{da_\bullet}{dt} = && (\dot{M} L_{\rm ms} - T_{\rm B})c/(GM_\bullet ^2) - \nonumber \\
&& 2 a_\bullet (\dot{M} c^2 E_{\rm ms} - \dot{E}_{\rm B}) /(M_\bullet c^2),
\end{eqnarray}
where $T_{\rm B}$ is the total magnetic torque applied on the BH, i.e.
\begin{eqnarray}
T_{\rm B} = \frac{\dot{E}_{\rm B}}{\Omega_{\rm F}}=3.4 \times 10^{45} a_\bullet^2 q^{-1} m^3 B_{\bullet,15}^2  F(a_\bullet) {\rm \ g \ cm^2 \ s^{-2}}. \nonumber \\
\end{eqnarray}

In the above Equations (\ref{dMbz}) and (\ref{dJbz}), we do not include the magnetic coupling effect between the BH and the disk through closed magnetic field lines (Li \& Paczynski 2000; Wang et al. 2002; Lei et al. 2009; Janiuk \& Yuan 2010). Similar to the Blandford-Znajek mechanism, the magnetic coupling effect also extracts rotational energy from the spinning BH. Only if the BH spin is initially small, the magnetic coupling would act as an additional spin-up process. A similar discussion on this aspect was made by Dai \& Liu (2012) within the context of the magnetar central engine model. In more general cases, the magnetic coupling effect would not significantly affect the BH spin evolution (Lei et al. 2009).

To delineate a GRB, we average the parameters over time. One may define $\bar\mu_{0} = \int \dot{E} dt /\int \dot{M}_{\rm j,BZ} c^2 dt$. We can calculate $\Gamma_0$ and $\Gamma_{\rm max}$ as discussed above. 

\begin{figure}[htc]
\center
\includegraphics[width=7cm,angle=0]{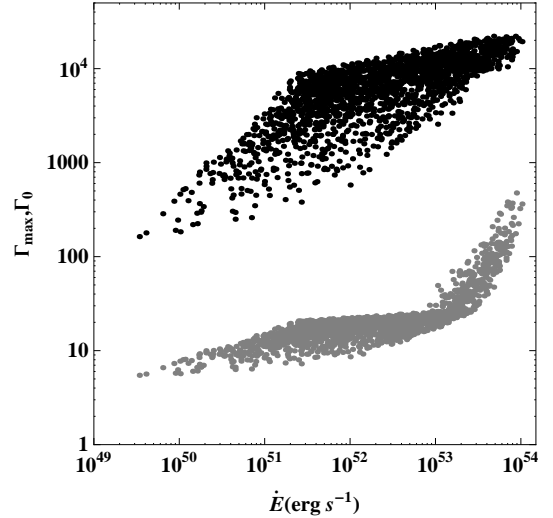}
\caption{$\Gamma_{\rm max}= \bar\mu_{0}$ (black) and $\Gamma_{0}$ (gray) vs. the central engine output power $\dot{E}$ within the framework of the BZ jet scenario. Two thousand GRBs with random BH mass, BH spin, accretion rate, and disk mass have been simulated. Following scaling correlations are found: $\Gamma_{\rm max} \propto \dot{E}^{0.32}$ and $\Gamma_0 \propto \dot{E}^{0.24}$. The Lorentz factor of the GRB during the prompt emission phase is expected to satisfy $\Gamma_0 < \Gamma < \Gamma_{\rm max}$.}
\label{fig2}
\end{figure}

In Fig.\ref{fig2}, we show the simulated 2000 GRBs in the same way as we did in Fig. \ref{fig1}. The same distributions for other parameters (BH spin, BH mass, accretion rate, disk mass, etc) have been adopted. Both $\Gamma_{\rm max}$ and $\Gamma_0$ have been plotted.

By fitting Fig.\ref{fig2}, we find the correlations $\Gamma_{\rm max} \propto \bar\mu_{0} \propto \dot{E}^{0.32}$ and $\Gamma_{0} \propto \dot{E}^{0.24}$. In the BZ model, the relativistic jet dissipates its magnetic energy via ICMART with an efficiency $\eta_{\rm ICMART}$ to produce gamma-ray emission, i.e., $L_{\gamma} \simeq \eta_{\rm ICMART} \dot{E}$. The efficiency can be approximated as (Zhang \& Yan 2011) $\eta_{\rm ICMART} \simeq 1/(1+\sigma_{\rm end})$, where $\sigma_{\rm end}$ is local magnetization parameter after the ICMART event. Typically one has $\eta_{\rm ICMART}>50\% $. So the observed $\gamma$-ray luminosity is a good proxy of the jet power $\dot E$. Again considering the beaming factor correction, one can derive 
 
$\Gamma_{\rm max} \propto \dot{E}^{0.30} \propto (f_b L_{\rm \gamma,iso})^{0.30} \propto L_{\rm \gamma,iso}^{0.26}$ and $\Gamma_0 \propto L_{\rm \gamma,iso}^{0.21}$. In reality, the real $\Gamma$ should satisfy eq.(\ref{gamma}). Since both limiting Lorentz factors show a correlation similar to the observations (Liang et al. 2010; L\"u et al. 2012), we suggest that the BZ model would also give a $\Gamma-L$ correlation that is generally consistent with the observations. Also the BZ jets are much cleaner than the neutrino-driven ones, which overcome the difficulty of the neutrino-driven jets that have too much baryon contamination in the jet. The relatively large values of $\Gamma_{\rm max}$ are not a big concern, since there could be other baryon loading processes besides the one considered here that would contaminate the jet even more.

\section{Conclusions and Discussion}

In this paper, we studied the baryon loading problem of a GRB jet launched by a hyper-accreting BH central engine. We considered two types of jet launching mechanisms: the non-magnetized $\nu\bar\nu$-annihilation mechanism and the strongly magnetized Blandford-Znajek mechanism. For both models, we considered baryons in a neutrino-driven wind from a hyperaccreting disk. For the $\nu\bar\nu$-annihilation model, the baryons launched in the neutrino-driven wind are mixed with the photons and electron-positron pairs produced by $\nu\bar\nu$-annihilation, and thermally accelerated to reach the termination Lorentz factor $\Gamma_{\rm max} \sim \eta$ (see definition in Equation [\ref{eta}]).  For a BZ jet, on the other hand, protons are blocked by the strong magnetic fields at the jet boundary. Only a fraction of neutrons can drift into the jet. We consider positron capture and proton-neutron inelastic collision processes and argue that about half of neutrons drifting into the jet can be converted to protons and be picked up by the jet. The other half of free neutrons would decay in the jet at larger radii, so that eventually all the neutrons can be added to the baryon loads of the jet. We calculated the minimum baryon loading of these magnetically dominated BZ jets in terms of the parameter $\mu_0$ (see definition in Equation [\ref{eq:mu}]). Since magnetic acceleration is inefficient, the final Lorentz factor of the GRB can be between $\Gamma_0$ (see definition in Equation [\ref{Gamma0}]) defined by initial efficient acceleration and $\Gamma_{\rm max} = \mu_0$.

A phenomenological correlation between GRB Lorentz factor measured in the deceleration phase and GRB isotropic gamma-ray luminosity has been discovered recently (Liang et al. 2010; L\"u et al. 2012), i.e. $\Gamma \sim L_{\rm iso}^{0.30}$. If GRB radiative efficiency does not sensitively depend on jet luminosity, this correlation would become a requirement for any GRB central engine model. With Monte Carlo simulations, we have shown that both baryon-loading models can give cleaner jets at high luminosities. The slope of dependence is consistent with the observations (see also L\"u et al. 2012 for a simpler $\nu\bar\nu$-annihilation model)\footnote{An alternative interpretation for this slope is based on the baryonic photosphere model (Fan et al. 2012). Even though the photosphere model may interpret the prompt emission spectrum of some special GRBs such as GRB 090902B (Ryde et al. 2010; Zhang et al. 2011; Pe'er et al. 2012), data analysis and theoretical modeling suggests that the ``Band'' function spectra observed in most GRBs are likely not of a photosphere origin (e.g. Zhang et al. 2011; Zhang et al. 2012; Guiriec et al. 2012). The observed $\Gamma-E_{\rm iso}$ and $\Gamma-L_{\rm iso}$ correlations must then stem from more fundamental central engine physics, as is discussed in this paper.}. The normalizations of the correlations are quite different for the two models. The $\nu\bar\nu$-annihilation jets are much dirtier than the BZ jets. For a typical jet opening angle $\theta_{\rm j} = \theta_{\nu\bar\nu} \sim 0.1$, we found that the resulting $\eta$ in the neutrino-driven jet is typically below a few hundreds, and is only a few 10s for typical GRB luminosities. These values are too small to be consistent with the GRB data. The BZ jets, on the other hand, are much cleaner. The $\Gamma_{\rm max}$ values are typically in the $10^3-10^4$ range. Since the ICMART mechanism (Zhang \& Yan 2011) can efficiently dissipate magnetic energy and prevent accelerating the jets to $\Gamma_{\rm max}$, and since there could be additional mechanisms to load more baryons in the magnetically dominated jets, we argue that the magnetically dominated central engine model is more appealing to interpret the GRB phenomenology. In view of the low normalization of the $\Gamma-L$ relation in the $\nu\bar\nu$-annihilation model, we suggest that at least a good fraction of GRBs should have a magnetically dominated central engine (see also M\'esz\'aros \& Rees 1997;  Wang et al. 2002; Lei et al. 2009; Yuan \& Zhang 2012; Fan et al. 2004; Fan et al. 2011).

In this paper, we did not consider the dependence of $\dot{M}_\nu$ on the strength of magnetic fields. In strong magnetic fields, electrons and positrons participating in the charged-particle reactions are restricted into discrete Landau levels, which may alter the neutrino heating and cooling rates, and therefore change $\dot{M}_\nu$. We will study these effects in future work. 

Besides the discussed two mechanisms to launch the jet, it is possible that an intrinsically episodic jet is launched from the disk through a magnetic process (Yuan \& Zhang 2012). The baryon loading process of this mechanism is more difficult to calculate, since baryons from the disk can be directly entrained in the magnetic bubble and escape. We do not discuss this mechanism in this paper. 

Overall, we restrict ourselves on the baryon loading problem in this paper. In a companion paper, we will study in detail how the BH central engine model may interpret the phenomenology of GRB prompt emission and X-ray afterglow.

\acknowledgements We thank other members of the UNLV GRB group and Shanqin Wang for helpful discussion and comments, and the anonymous referee for helpful suggestions. This work is supported by NSF under Grant No. AST-0908362, by National Natural Science Foundation of China (grants 11003004, 11025313, 11173011 and U1231101), and National Basic Research Program (``973'' Program) of China under Grant No. 2009CB824800. WHL acknowledges a Fellowship from China Scholarship Program for support. BZ acknowledges the UNLV sabbatical review committee and a Cheung Kong Scholar fellowship at Peking University in China.





\end{document}